\documentclass[twocolumn,aps,floatfix]{revtex4}
\usepackage{longtable}
\usepackage{amsmath,bm}%
\usepackage{graphicx}%
\usepackage{epsf,epsfig,latexsym}
\begin{document}
\title{A search for effects of  short range correlations in proton spectra
              from collisions induced by 47 MeV/u projectiles.
}

\author{K. Hagel}
\affiliation{Cyclotron Institute, Texas A$\&$M University, College Station, Texas 77843, USA}
\author{J. B. Natowitz}
\affiliation{Cyclotron Institute, Texas A$\&$M University, College Station, Texas 77843, USA}

\date{\today}

\begin{abstract}
An analysis of the energy spectra of protons emitted in reactions of 47 MeV/u projectiles with Sn and Au targets provides evidence for high momentum tails in the intrinsic momenta spectra of the projectiles. These high momentum spectra decrease with wavenumber $k$ at a rate proportional to $1/k^4$.  We suggest that additional experiments could provide a more refined value for the value of the power law exponent.  
\end{abstract}



\maketitle

\section{Introduction}
Theoretical studies of nuclei have long supported the existence of tensor force induced short range correlations in nuclei. Such correlations were predicted to lead to high momentum tails in the nucleon momentum distributions, extending far beyond the Fermi momentum [1-9]. While some indications of such high momentum nucleons appeared in various early experimental studies employing light projectiles [10,11], this phenomenon has, to date,  received very little attention from experimentalists working in the area of fermi energy heavy ion collisions and in general is not incorporated into modern transport model calculations which are routinely employed in efforts to derive nuclear equation of state information from comparisons to experimental data [12 and references therein]. This lack of attention appears to reflect both a commonly held belief that rapid thermalization of the nucleon distribution during the collisions will eclipse the effects of the short range correlations as well as the difficulties encountered in treating these correlations within mean field theoretical models. However, it is precisely because both nucleon-nucleon collisions and mean-field effects are important in the evolution of near fermi energy heavy ion collisions that much effort has been devoted to understanding the nature of the collision region and the extent to which equilibration of various degrees of freedom, thermal, chemical, N/Z ratio, etc is realized. 

A widely accepted picture of the reaction process is that, at the early stage of the collision, a compressional and thermal shock creates a hot composite nucleus that may expand to low densities and form clusters. Since the light particle emission, which occurs during the collisions, witnesses each stage of the reaction, it carries essential information on the early dynamics and on the degree of equilibration at each stage [13-15]. The kinematic features and yields of emitted light particles and clusters can exploited to probe the nature of the evolving system and information on the EOS can be extracted. An initial estimation of particle and cluster emission multiplicities is often made by fitting the observed spectra assuming contributions from three sources, a projectile-like fragment (PLF) source, an intermediate velocity (IV) source, and a target-like fragment (TLF) source. A reasonable reproduction of the observed spectra is achieved. The IV source velocities are very close to 50\% of the beam velocity indicating its generation by nucleon-nucleon collisions. Investigations of the dynamics in a large series of heavy ion reaction experiments indicate that much of the early particle emission may be attributed to such nucleon-nucleon collisions [13-15].  However, only in recent years has an appreciation that the high momentum tails could prove to be significant in nuclear equation of state studies triggered renewed interest in investigating such effects. This renewed interest was spurred by results of electron scattering studies [16-20], the most recent of which  demonstrate that $\sim20$\% of nucleons in the nucleus have momenta higher than the Fermi momentum. 

While features reflecting such short range correlations may not be easily discerned in all reaction observables, there might be significant effects on the observables reflecting the earliest stages of the reaction. Since experiments focusing on such observables have become increasingly important in laboratory efforts to probe the nuclear equation of state at higher than normal nuclear densities [21,22] careful consideration should be paid to the role of short range correlations and their proper incorporation into transport models. Indeed in a recent theoretical work it is suggested that the high momentum tail in the momentum distribution has a significant effect on early photon emission and thus must be taken into account in extracting valid EOS parameters, e.g. the symmetry energy at supra normal density, from the observed cross sections and angular distributions [21]. It is further argued that, although the commonly predicted slope of the high momentum tail is 1/$k^4$ [1-9]  such photon emission experiments could serve to more precisely establish the spectral shape of the high momentum tail, providing a sensitive test of the short range interaction.

\begin{figure}
\epsfig{file=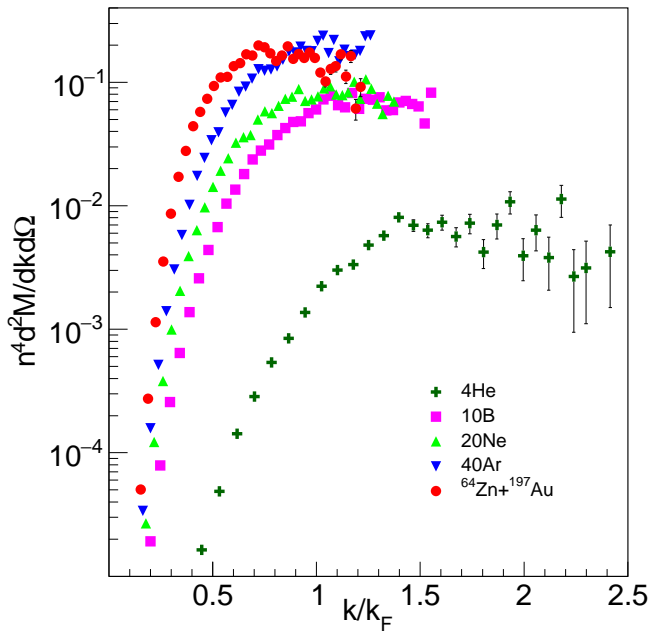,width=9.2cm,angle=0}
 \caption{Differential cross sections d$\sigma$/d$k$d$\Omega$ for proton emission at $\theta_{lab} = 4.2^o$ multiplied by $k^4$  and plotted against $k/k_F$(A). See text.  }
\label{fig1}
\end{figure}

\section{Analysis and results}
Based upon the above information it is natural to ask whether the nucleon spectra, themselves, manifest the effects of high momentum tails. In previously reported experiments our group employed the NIMROD $4\pi$ multi-detector at Texas A$\&$M University [23] to study light particle and cluster emission from  collisions of 47 MeV/u $^4$He, $^{10}$B, $^{20}$Ne, $^{40}$Ar and $^{64}$Zn on $^{112, 124}$Sn [13,14] and 47 MeV/u, $^{64}$Zn with $^{197}$Au [13-15]. Further details on the detection system, energy calibrations and neutron ball efficiency may be found in [13-15,23].  We have used these data to search for evidence of high momentum tails in the proton spectra measured in these reactions.

The differential multiplicity energy spectra from spectra obtained in the studied reactions were analyzed by first performing a relativistic transformation of the very small laboratory angle ($\sim4.2$ degrees) proton spectra into the frame of the projectile. The resultant momenta spectra were converted  into wavenumber ($k = p/\hbar$), These spectra were subjected to a small free path correction for transmission of nucleons through the target nucleus [24]. This correction has a very small effect on the spectral shape at high momentum. The spectrum  d$\sigma$/d$k$d$\Omega$ was multiplied by $k^4$. The resulting distributions are plotted in Figure 1 as a function of $k/k_F$(A) where  $k_F$(A) for each projectile mass, A, has been derived from interpolation of the experimental results of reference [25] We take the essentially flat horizontal distributions observed at high momenta  to confirm the existence of high momentum tails having a $1/k^4$  dependence.  This, plus the progressive increase in cross section with increasing projectile mass for these particles appearing at the highest momenta in the small angle spectra suggests that they are very early emitted Fermi jet like particles which have not suffered nucleon-nucleon collisions[26,27].

\begin{figure}
\epsfig{file=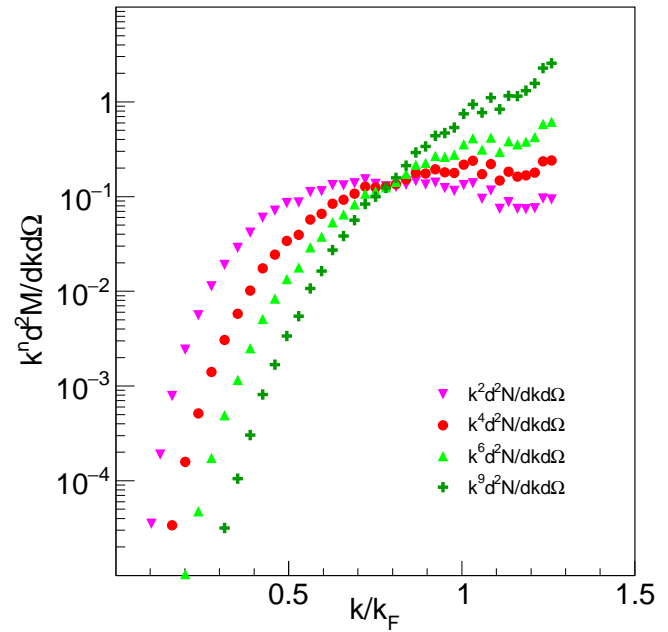,width=9.2cm,angle=0}
 \caption{Figure 2. A test of alternative exponents for the power law slope of the proton spectrum for the 47 MeV/u induced reaction at high momenta. See text.}
\label{fig2}
\end{figure}

As a further investigation of the slope factor we present in Figure 2 the differential cross section d$\sigma$/d$k$d$\Omega$ for $^{40}$Ar + $^{112,124}$Sn multiplied by $k$ raised to various powers, i.e., 2, 4, 6, and 9 [21]. The data appear to rule out 2, 6 and 9. It is possible however that systematic additional experiments at different energies and with better statistics might provide a more refined value of the power law exponent. 

\section{Acknowledgments}
This work was supported by the United States Department of Energy under Grant No. DE-FG03-93ER40773.

\end{document}